\documentstyle[aps,multicol,epsf]{revtex}
\begin{document}
\newcommand{\beq}{\begin{eqnarray}}
\newcommand{\eeq}{\end{eqnarray}}
\newcommand{\bk}{{\bf k}}
\title{Minimum Thermal Conductivity of Superlattices}
\draft
\author{M.V. Simkin and G.D. Mahan}
\address{Department of Physics and Astronomy, University of Tennessee, 
Knoxville, 37996-1200, and\\
Solid State Division, Oak Ridge National Laboratory, P.O. Box 2008, 
Oak Ridge, TN, 37831}
\date{\today}
\maketitle
\begin{abstract}
The phonon thermal conductivity of a multilayer is calculated for transport 
perpendicular to the layers. There is a cross over between particle transport 
for thick layers to wave  transport for thin layers. The calculations 
shows that the conductivity has a minimum value for a layer thickness 
somewhat smaller then the mean free path of the phonons.
\end{abstract}
\begin{multicols}{2}
\narrowtext
The thermal conductivity is a fundamental transport parameter\cite{zim}. There has been much
recent interest in the thermal conductivity of semiconductor superlattices due to
their possible applications in a variety of devices. Efficient solid state
refrigeration requires a low thermal conductivity\cite{gdm98}.  Preliminary experimental and
theoretical work suggests that the thermal conductivity of superlattices is quite low,
both for transport along the planes\cite{yao,hyl2}, or perpendicular to the
planes\cite{mar1,lee,chen,hyl1}. The heat is carried by excitations such as phonons and
electrons. Most theories use a Boltzmann equation which treats the excitations as particles 
and ignores wave interference\cite{chen,chen2}. These theories all predict that the
thermal conductivity perpendicular to the layers decreases as the layer spacing is reduced
in the superlattice. The correct description using the Boltmann equation would be to
use the phonon states of the superlattice as an input to the scattering, but this has not yet
been done by anyone.

We present calculations of the thermal conductivity perpendicular to the layers which
includes the wave interference of the superlattice. These calculations, in one,
two, and three dimensions always predict that the thermal conductivity {\it
increases} as the layer spacing is reduced in the superlattice. This behavior is
shown to be caused by band folding in the superlattice. It is a general feature which
should be true in all cases. The particle and wave calculations are in direct
disagreement on the behavior of the thermal conductivity with decreasing layer
spacing. This disagreement is resolved by calculations which include the
mean-free-path (mfp) of the phonons. For layers thinner than the mfp, the wave
theory applies. For layers thicker than the mfp the particle theory applies. The
combined theory predicts a minimum in the thermal conductivity, as a function of layer
spacing. The thickness of the layers for minimum thermal conductivity depends upon
the average mfp, and is therefore temperature dependent.

The particle theories use the interface boundary resistance\cite{poh} as the important
feature of a superlattice. A superlattice with alternating layers has a thermal resistance
for one repeat unit of $R_{SL} = L_1/K_1+L_2/K_2 +2R_B$, where $(L_j,K_j)$ are the
thickness and thermal conductivity of the individual layers, and $R_B$ is the thermal
boundary resistance. For simplicity assume that $L_1=L_2\equiv L$, which is often the case
experimentally. The effective thermal conductivity of the superlattice is then
\beq
K_{SL}&=& \frac{2L}{R_{SL}} = \frac{2L}{L(1/K_1+1/K_2)+2R_B}
\eeq
This classical prediction is that the thermal conductivity decreases as the layer
thickness $L$ decreases\cite{chen2}. 

The wave theory calculates the actual phonon modes $\omega_{\lambda}(\bk)$ of the
superlattice, where $\lambda$ is the band index. They are used to calculate the thermal
conductivity from the usual formula in $d$-dimensions\cite{zim},
\beq
K(T)&=& \sum_{\lambda}\int \frac{d^dk}{(2\pi)^d}
\hbar\omega_{\lambda}(\bk)|v_z(\bk)|\ell_{\lambda}(\bk)\frac{\partial n(\omega,T)}{\partial T}
\eeq
where $n(\omega,T)$ is the Bose-Einstein
distribution function.   A rigorous treatment uses Boltzmann theory applied to the transport in
minibands to find the mean-free-path $\ell_{\lambda}(\bk)$. 
 At high temperatures, one can approximate $n \sim
k_BT/\hbar\omega_{\lambda}(\bk)$ which gives the simpler formula
\beq
K(T)&=& k_B\sum_{\lambda}\int \frac{d^dk}{(2\pi)^d}|v_z(\bk)|\ell_{\lambda}(\bk)
\eeq
The above formula is quite general. There are two important special cases of constant
relaxation time $(K_{\tau})$ and constant mfp $(K_{\ell})$
\beq
K_{\tau}(T)&=& k_B\tau\sum_{\lambda}\int \frac{d^dk}{(2\pi)^d}v_z(\bk)^2\\
K_{\ell}(T)&=& k_B\ell\sum_{\lambda}\int \frac{d^dk}{(2\pi)^d}|v_z(\bk)|
\eeq 
Both of these formulas can be related to the distribution $P(v_z)$ of phonon
velocities perpendicular to the layers
\beq
P(v_z)&=& \sum_{\lambda}\int \frac{d^dk}{(2\pi)^d}\delta(v_z-|v_z(\bk)|)\\
K_{\tau}&=& k_B\tau \int d v_z P(v_z)v_z^2\\
K_{\ell}&=& k_B\ell \int d v_z P(v_z)v_z.
\eeq
Wave interference leads to band folding\cite{dow,rid}. Band folding leads to a reduction
of the phonon velocities. Both $K_{\tau}$ and $K_{\ell}$ are reduced by wave
interference. The case of constant mfp in one dimension can be reduced to a 
simple formula 
\beq
K_{\ell}&=& \frac{k_B\ell}{2\pi}\sum_{\lambda}\int dk|\frac{d\omega}{dk}|\\
&=&
\frac{k_B\ell}{2\pi}\sum_{\lambda}[\omega_{\lambda}(k_{max})-\omega_{\lambda}(k_{min})]\\
&=& \frac{k_B\ell}{2\pi}[\omega_{max}-(\mbox{energy gaps})]
\eeq
The integration is eliminated.

 These assertions are best illustrated in one dimension, that is with
the atomic chain. The  simplest model for a superlattice has all spring 
constants
identical, and the layers differ in their masses. Layer one has $N/2$
atoms of mass $m_1$ and layer two has $N/2$ atoms of mass $m_2=m_1/\alpha$. The
characteristic matrix for phonons is given below for the case $N=6$
\beq
{\bf M} = &&\left|\begin{array}{cccccc}
2 & -1 & 0 & 0 & 0 & -e^{-ikN}\\
-1 & 2 & -1 & 0 & 0 & 0 \\
0 & -1 & 2 & -1 & 0 & 0 \\
0 & 0 & -\alpha & 2\alpha & -\alpha & 0 \\
0 & 0 & 0 & -\alpha & 2\alpha & -\alpha \\
-\alpha e^{ikN} & 0 & 0 & 0 & -\alpha & 2\alpha \end{array}\right|\label{cm}
\eeq
Setting the determinant $||{\bf M}-\omega^2 {\bf 1}||$ to zero gives the 
characteristic equation (the easiest way to derive it is to utilize the 
similarity of the present problem to the Schr\"{o}edinger equation with the
Kronig-Penney potential) $k(\omega)$ of \cite{col,ryt,jus}
\beq
&&\cos(kN) = \cos(k_1N/2)\cos(k_2N/2)-\nonumber\\
&&\frac{1-\cos(k_1)\cos(k_2)}{\sin(k_1)\sin(k_2)}\sin(k_1N/2)\sin(k_2N/2) \label{disp}
\eeq
where $\cos(k_1)=1 - \omega^2/2, \cos(k_2)=1-\omega^2/(2\alpha)$ define the 
wave
vectors $(k_1,k_2)$ in the individual layers in dimensionless units. A typical spectrum
is shown in Fig.1, where $\alpha =2$ and the superlattice periods are $N=2,4,8,16$. Modes with
frequency $\omega>2$ are confined within the layer of the lighter atom, and contribute
little to the thermal conduction. As the value of $N$ is increased in Fig. 1, there is
more band folding, and the average velocity decreases. The lower curve in Fig.2 shows the thermal
conductivity as a function of superlattice period. The result for constant mfp is normalized
to $k_B\ell \omega_{2,\mbox{max}}$, where $\omega_{2,\mbox{max}}=2$ is
the maximum phonon frequency in the layer 1 of heavy mass. The heat
conduction is highest at small values of $N$, and rapidly decreases as $N$ increases.
 It also shows an irregular nonmonotonous behavior which we understand and 
will be discussed elsevere. This result is for one dimension. Similar curves are found for every case
which we have calculated: for different values of mass ratio $\alpha$, and for both $K_{\ell}$
and $K_{\tau}$. Generally, increasing $N$: (i) increases the amount of band folding,
(ii) decreases the average velocity in the superlattice, and (iii) decreases the thermal
conductivity.

Lower curves in Figs.3 and 4 show similar calculations of $K_{\ell}$ for mass ratio $\alpha=2$ in two and
three dimensions. Including only nearest-neighbor interactions, the characteristic
matrix (\ref{cm}) is changed only in its diagonal element, where 2 is replaced by
$2[2-\cos(q)]$ in two dimensions, and $2[3-\cos(q_x)-\cos(q_y)]$ in three dimensions.
Here $(q,q_x,q_y)$ are the wave vectors within the planes. These variables are
integrated to find the result for the heat conduction.  These cases also have the
feature the thermal conducivity falls with increasing value of layer thickness. The
curve is now smooth, due to the averaging over the parallel wave vectors. Similar
results are found for all values of $\alpha$.

 In these
calculations the thermal boundary resistance seems to have disappeared. Note that our
model does predict thermal boundary resistance for a single interface\cite{you}. The
scattering of phonon waves at the interface causes the thermal resistance.  This scattering
is included in the present calculation. However, it also causes band folding, which is a
bigger effect than the thermal boundary resistance. The wave calculation predicts
that in three dimensions the thermal conductivity increases as the layer spacing is
decreased, which is exactly the opposite of the particle calculation which predicts
that it decreases. Since the effect is due simply to band folding, which is a well
documented phenomena, then the prediction must be accurate. Several prior calculations
predicted behavior similar to that shown in figures 2,3,4 \cite{hyl3,mar2}. 
However, no explanation was given for the behavior. 

The missing ingredient in these calculations is the mean-free-path  of the
phonons. When the layer thickness exceeds the mfp, then interference effects should
diminish, and the particle model should become applicable. Our intuition is that the
wave model should apply when $L<\ell$ and the particle model should apply when
$L>\ell$. An phenomenological  method of including $\ell$ is to add a complex part to the wave
vector $k$ which is $i/\ell$.
Then recalculate the properties of the
superlattice using eqn.(\ref{disp}). 
This idea came from Pendry\cite{pen} who
did the same thing for electron energy bands (and provides a more complete
justification for this model): energy gaps go away if one 
includes a finite mfp. A similar result is found for phonons. The band 
gaps diminishe to zero as the mfp
is decreased for a fixed value of $N$.

Upper curves in Figs. 2,3,4 show the thermal conductivity in one, two and three dimensions as a function of superlattice period for four
different values of mfp and $\alpha =2$. The mfp is given in terms of the number of
lattice spacings (note: not superlattice spacings).  Similar curves are found for other values of
$\alpha$. For large values of $\ell$ the results are identical to the lower curve (which has
$\ell=\infty$). For small values of mfp $(\ell=10$) the thermal
conductivity is nearly independent of superlattice period, except where it falls at small values of
$N$. 


The interesting cases have $\ell=50$ or $100$. Here the thermal conductivity falls as
$N$ increases, reaches a minimum, and then starts to increase. This latter behavior is the
situation expected in the experiments. At room temperature, in most solids, anharmonic
scattering limits the phonon mpf to value in the range of 10-100 lattice constants, which
is also the typical value of superlattice parameter in current devices. Therefore we expect the
experimental thermal conductivities to behave as the curves marked $\ell=50,100$ in Fig.2,3,4. The
thermal conductivity should have a minimum value when plotted vs. superlattice period. The minimum
occurs at the cross over between the particle and wave-interference types of
transport. One experimental result  has this behavior\cite{ven}.

We thank P. Hyldgaard for useful conversations. 
Research support is acknowledge from the Office of Naval Research Contract No.
N00014-98-1-0742, from the University of Tennesee, and   from Oak Ridge National Laboratory
managed by Lockheed Martin Energy Research Corp. for the U.S. Department of Energy under
contract DE-AC05-96OR22464.  

%
%
 \begin{figure}
\centering
\begin{minipage}{8.0cm}
\epsfxsize= 8 cm \epsfbox{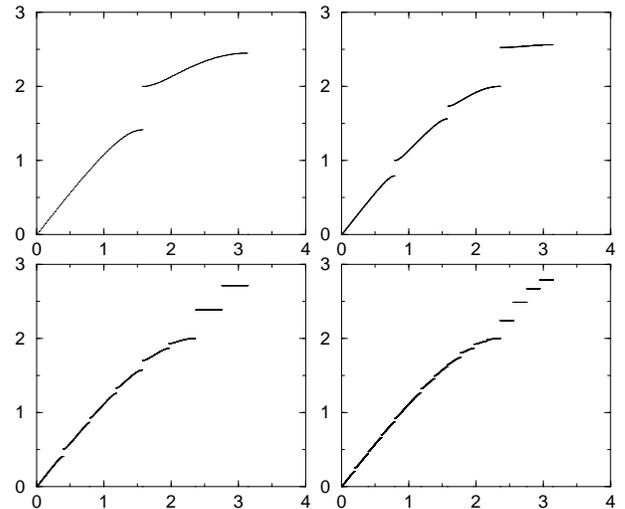}
\end{minipage}
\caption{Spectrum of $N=2,4,8,16$ superlattices with mass ratio $\alpha=2$ in the
extended zone representation.}
 \end{figure}
\begin{figure}
\centering
\begin{minipage}{8.0cm}
\epsfxsize= 8 cm \epsfbox{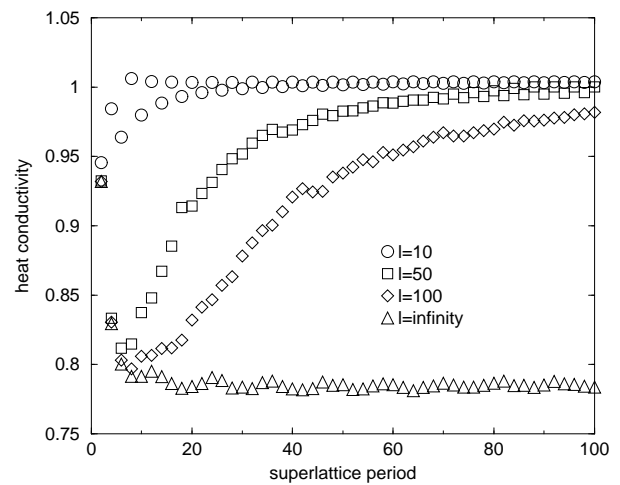}
\end{minipage}
 \caption{Heat conductivity in one dimension as a function of superlattice period for mass
ratio $\alpha =2$ for different values of the phonon mfp which is given in units of lattice periods. Dimensionless units found by dividing (11) by $2k_B\ell$.}
 \end{figure}
\begin{figure}
\centering
\begin{minipage}{8.0cm}
\epsfxsize= 8 cm \epsfbox{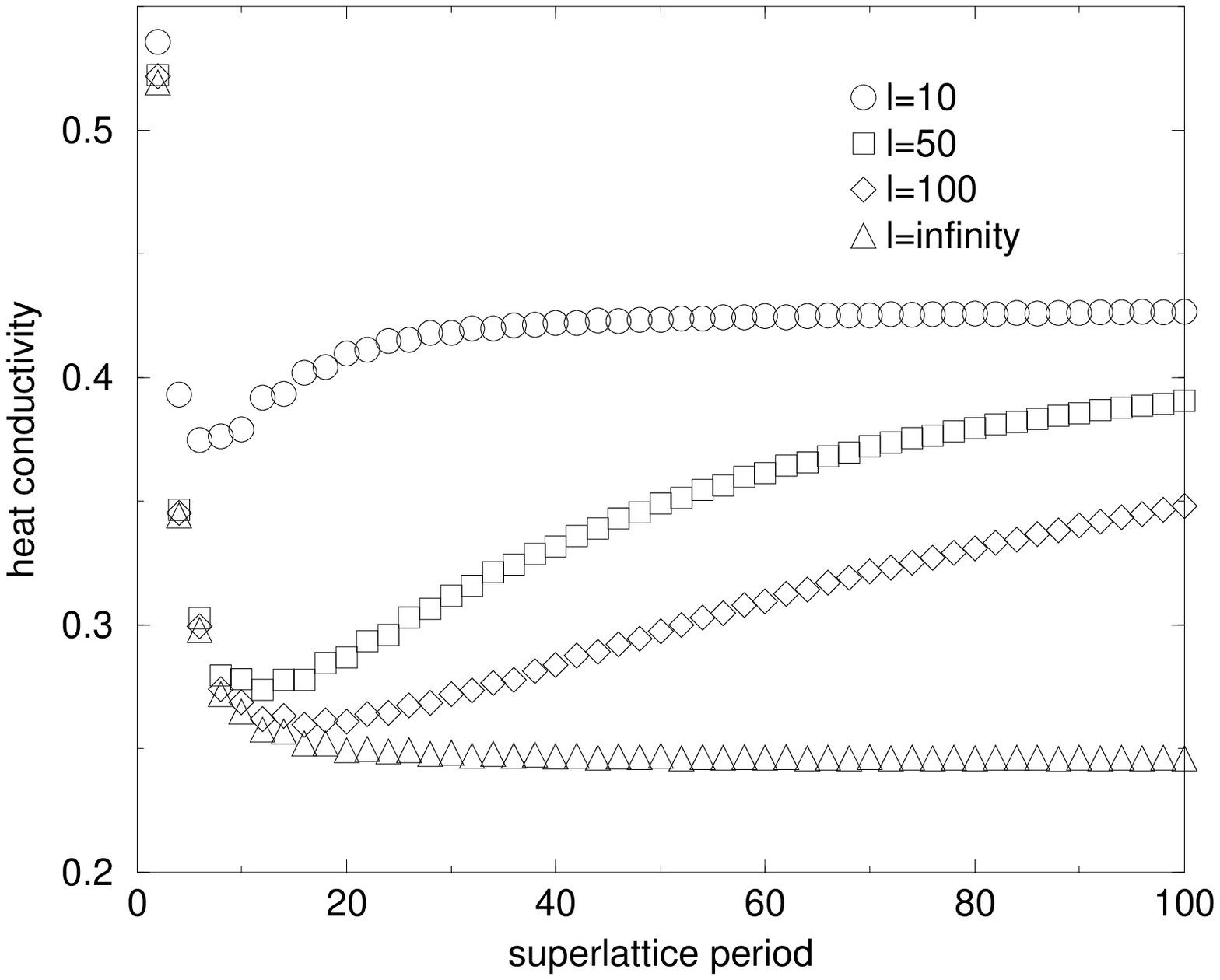}
\end{minipage}
\caption{Thermal conductivity in 2 dimensions as a function of superlattice
period for mass ratio $\alpha=2$ for different values of the phonon mfp which is given in units of lattice
periods. (In units of thermal conductivity of a uniform system of the
heavier atoms.)}
 \end{figure}
\begin{figure}
\centering
\begin{minipage}{8.0cm}
\epsfxsize= 8 cm \epsfbox{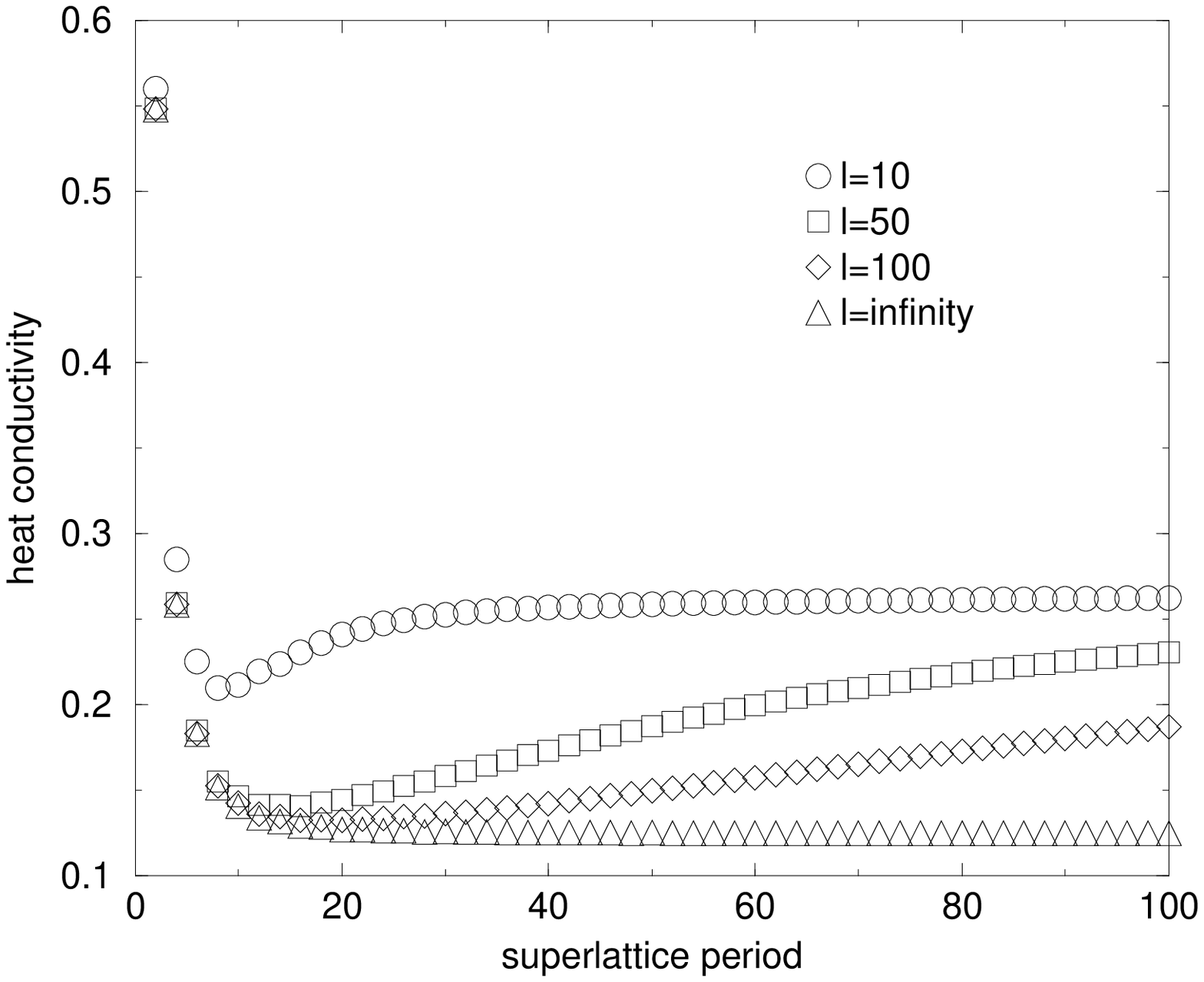}
\end{minipage}
\caption{Thermal conductivity in 3 dimensions as a function of superlattice
period for mass ratio $\alpha=2$ for different values of the phonon mfp which is given in units of lattice
periods. (In units of thermal conductivity of a uniform system of the
heavier atoms.)}
 \end{figure}
%
%
\end{multicols}
\end{document}